\documentclass[reprint,amsmath,amssymb,prl,superscriptaddress]{revtex4-2}
\usepackage{graphicx}
\newcommand\vs[1]{\color{red}}
\usepackage{color}
\usepackage{soul}
\begin{document}
\title{Elastic instability in a straight channel of viscoelastic flow without prearranged perturbations}

\author{Yuke Li}
\affiliation{Department of Physics of Complex Systems,
Weizmann Institute of Science, Rehovot 7610001, Israel}
\author{Victor Steinberg}
\affiliation{Department of Physics of Complex Systems,
Weizmann Institute of Science, Rehovot 7610001, Israel}
\affiliation{The Racah Institute of Physics, Hebrew University of Jerusalem, Jerusalem 91904, Israel}

\date{\today}

\begin{abstract}
We report experimental results on elastic instability in a viscoelastic channel shear flow due to only a natural non-smoothed inlet and small holes along the channel for pressure measurements. We show that non-normal mode instability results in elastic waves and chaotic flow self-organized into periodically cycled stream-wise streaks synchronized by elastic wave frequency. The chaotic flow persists above the transition with increasing $Wi$ further into elastic turbulence and drag reduction regimes. Thus, we resolve the recent puzzle whether strong prearranged perturbations are necessary to get an elastic instability in parallel shear viscoelastic flow. Moreover, flow resistance, velocity spectra decay, and elastic wave speed reveal the same scaling with $Wi$ as obtained in the case of strong disturbances. The remarkable result is that the all scaling behavior and streaks are found in the entire channel with small attenuation, in a sharp contrast to the flow with strong prearranged perturbations.
\end{abstract}

\maketitle

In parallel shear Newtonian flows, turbulence emerges at the high Reynolds numbers ($Re$) due to inertial stresses, whereas for low $Re$, they are dominated by viscous dissipation and remain laminar. In contrast, in viscoelastic fluid flows with curved streamlines, at $Re\ll1$ and the high Weissenberg number, $Wi$, a linear elastic instability due to a single fastest growing mode results in a single most stable mode, which leads to elastic turbulence (ET) at vanishing inertia, as $Wi$ is increased further. ET is an inertia-less, chaotic flow driven purely by elastic stress generated by polymers stretched by the flow. The latter is modified by a feedback reaction of the elastic stress. As the result, the elastic "hoop" stress, generated by polymers stretched along the curved streamlines, engenders a force towards the curvature center, which triggers an elastic instability \cite{Larson_JFM1990,shaqfeh1996purely} and ET at $Wi\gg1$ and $Re\ll1$ \cite{Fouxon_PoF2003,Victor_NatureET2000,Victor_review_2021}. The elastic instability mechanism becomes ineffective in zero curvature limit \cite{Larson_JFM1990,pakdel1996elastic}, in such as parallel shear flows, where their linear stability in the whole range of $Wi$ is proved \cite{Leonov_JMM1967,Renardy_JNFM1986}, similar to Newtonian parallel shear flows. Here $Wi=\lambda U/h$ defines the degree of polymer stretching \cite{BirdBook}, $Re=\rho Uh/\eta$, where $U$ is the mean flow velocity, $\lambda$ is the longest polymer relaxation time, $\rho, \eta$ are the solution density and dynamic viscosity, respectively, and $h$ is the characteristic vessel scale.

Parallel shear flows of viscoelastic fluids are proved to be linearly stable at all $Wi$ \cite{Leonov_JMM1967,Renardy_JNFM1986}, though it does not imply their global stability. Indeed, two recent theoretical developments \cite{morozov2005subcritical,hoda2008energy,jovanovic2011nonmodal,lieu2013worst,page2014streak} suggest two different possible mechanisms of instability of parallel shear viscoelastic flows. In spite of different nature of these mechanisms, namely normal versus non-normal unstable modes, both mechanisms consider finite-size perturbations as a necessary condition for the instability. First mechanism is based on a hysteretic, sub-critical finite-amplitude bifurcation, based on a nonlinear extension of the normal mode linear stability analysis, predicts a single normal mode, namely 2D traveling waves, growing exponentially and saturated at a sufficiently large amplitude, which at higher $Wi$ may result in ET  \cite{morozov2005subcritical}. The second approach utilizes theoretically and verified experimentally a non-modal analysis of linearly stable shear Newtonian flows at $Re\gg1$ \cite{schmid2007nonmodal,trefethen1993hydrodynamic,waleffe1997self,schoppa2002coherent}. This approach used in theory and numerical simulations for the non-modal instability of viscoelastic channel and plane Couette flows at $Wi\gg1$ and $Re\ll1$ (and the elasticity number $El=Wi/Re\gg1$) is introduced in Refs. \cite{hoda2008energy,jovanovic2011nonmodal,lieu2013worst}. The resulting span-wise modulated coherent structures are revealed in plane Couette and channel flows of visco-elastic fluids \cite{hoda2008energy,jovanovic2011nonmodal,lieu2013worst,page2014streak}. The elastic instability in viscoelastic parallel shear flows is considered to occur via transient stream-wise streaks generated by weakly unstable stream-wise vortices \cite{lieu2013worst,page2014streak}. Both coherent structures (CSs) are the most amplified, similar to those observed in Newtonian shear flows at $Re\gg1$ \cite{schoppa2002coherent,waleffe1997self}. Since the theory is based on the linearized equations for the non-normal pseudo-modes, it describes only the linear transient stages of the instability and cannot predict CSs prevailed in at saturation due to nonlinear interactions. In the present study, we show how the nonlinear interactions stabilize CSs selected by the linearized dynamics and eventually promote transition to ET at higher $Wi$.


Recent experiments in pipe \cite{bonn2011large} and straight square micro-channel \cite{pan2013nonlinear,qin2017characterizing,qin2019flow} with large prearranged perturbations, generated either by the jet \cite{bonn2011large} or the obstacle array located at the inlet \cite{pan2013nonlinear}, reveal an elastic instability directly into chaotic flow characterized by continuous velocity and pressure power spectra. Moreover, the authors of Ref. \cite{pan2013nonlinear} claim observing a hysteresis in the transition suggested in Ref. \cite{morozov2005subcritical}. However, the predicted backward instability results in the most unstable normal mode that is a 2D traveling wave, which is not detected in the experiment.
These experimental results suggest that the elastic instability and probably ET may occur in parallel shear viscoelastic flows, through a direct transition to a chaotic states with a large number of the excited modes observed experimentally in both parallel shear viscoelastic flows inconsistent with a normal mode instability  \cite{bonn2011large,pan2013nonlinear,qin2017characterizing}.

Recently, a non-modal instability in a parallel shear channel flow is confirmed in a straight channel with strong perturbations at the inlet \cite{jha2020universal,jha2021elastically}, as suggested in Ref.  \cite{hoda2008energy,jovanovic2011nonmodal,lieu2013worst,page2014streak}. Furthermore, elastic waves and chaotic flow, observed only in a part of the channel, adjacent to the perturbation source, are self-organized into regularly cycled stream-wise vortices and streaks synchronized by elastic wave frequency in all three flows regimes \cite{jha2020universal}. 
Throughout the rest of a channel flow, elastic waves and CSs decay, with normalized perturbation intensity that reduces from roughly $10\%$ down to $4\%$ downstream until the outlet \cite{jha2020universal}.

 
In this Letter, we present results that answer two questions, namely: (i) Whether three different chaotic flow regimes, occurred above an elastic instability as the result of finite-size perturbations due to a non-smoothed inlet and six small holes, exhibit the same scaling of the flow properties and elastic wave speed with $Wi$, as observed in the flow with strong prearranged perturbations \cite{jha2020universal,jha2021elastically}; and (ii) Either these chaotic flows are self-organized into periodically cycled CSs, stream-wise rolls and streaks synchronized by elastic wave frequency or only into stream-wise streaks, similar to Ref. \cite{shnapp2021nonmodal}. The key observation is that the scaling behavior in three flow regimes and streaks are found in the entire channel with small attenuation, in a sharp contrast to Refs. \cite{jha2020universal,jha2021elastically}.


\begin{figure}
\includegraphics{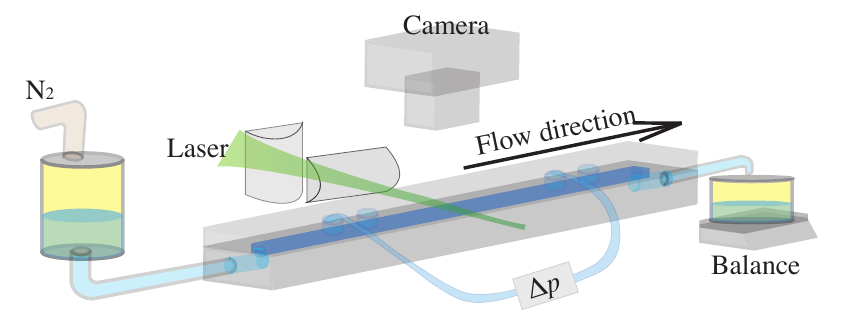}
\caption{
Schematics of experimental setup. Four instead of six holes are drawn for simplicity.}


\label{fig:1_schematics}
\end{figure}

The experiments are conducted in a straight long channel, with dimensions 500(\textit{L}) $\times$ 3.5(\textit{w}) $\times$ 0.5(\textit{h}) mm$^3$, shown in Fig.\ref{fig:1_schematics}. The only main possible source of flow disturbances is the channel's non-smoothed inlet and six holes in the top plate for pressure measurements.
As the working fluid, we use the polymer solution prepared viscous aqueous solvent with 64\% sucrose with dissolved high-molecule-weight Polyacrylamide (PAAm, Polysciences, $M_w=18 MDa$) at a dilute concentration of 80ppm. The solution properties are the solution density $\rho=1320$ kg/m$^3$, the solvent and solution viscosity $\eta_s=0.13$ Pa$\cdot$s and $\eta = \eta_s+\eta_p=0.17$ Pa$\cdot$s, respectively, $\eta_p/(\eta_s+\eta_p)\approx0.3$, where  $\eta_p$ is the polymer contribution into the solution viscosity, and the longest polymer relaxation time $\lambda=13$ s using the stress relaxation method \cite{SolutionPreparation}. 


The fluid in the channel is driven by N$_2$ gas by pressurized up to 100 psi. During experiments, we use a PC-interfaced balance (BPS-1000-C2-V2, MRC) to measure the time-averaged  fluid mass discharge rate at the channel exit, $Q=\langle\Delta m/\Delta t\rangle$, where $m(t)$ is weight instantaneously fluid as a function of time.
Then the mean velocity is calculated as $U=Q/\rho w h$, and $\mathrm{Wi}=\lambda U/ h$ and $\mathrm{Re}=\rho Uh/\eta$, which vary in the ranges (30, 6000) and (0.005, 0.9), respectively. We also measure pressure drops for flow resistance and fluctuations using high resolution differential pressure sensors of accuracy 
in various ranges: 5, 30, and 60 psi (HSC series, Honeywell).

We conduct measurements of the velocity field at various distances $l/h$ downstream from the inlet, using the particle image velocimetry (PIV) method. For that we illuminate small latex particles (3.2$\mu$m fluorescent tracers with concentration $\sim$0.67\% w/w, Latex Microsphere, Thermo Scientific) by a laser sheet with $\sim 100 \mu$m thickness over the middle plane in the channel. Then we capture pairs of images of the tracers using high-speed (frame rate from 500 up to 8000) and high spatial resolution (up to 2048$\times$2048 px$^2$) camera (Mini WX100, Photron  FASTCAM). The OpenPIV software \cite{OpenPIV} is employed to analyze $u(x,z,t)$ and $w(x,z,t)$ in 2D $x$-$z$ plane. We typically record data for periods of $\sim\mathcal{O}(15)$ minutes or $\sim\mathcal{O}(50\lambda)$ for each $Wi$ to obtain sufficient statistics.

\begin{figure}
\includegraphics{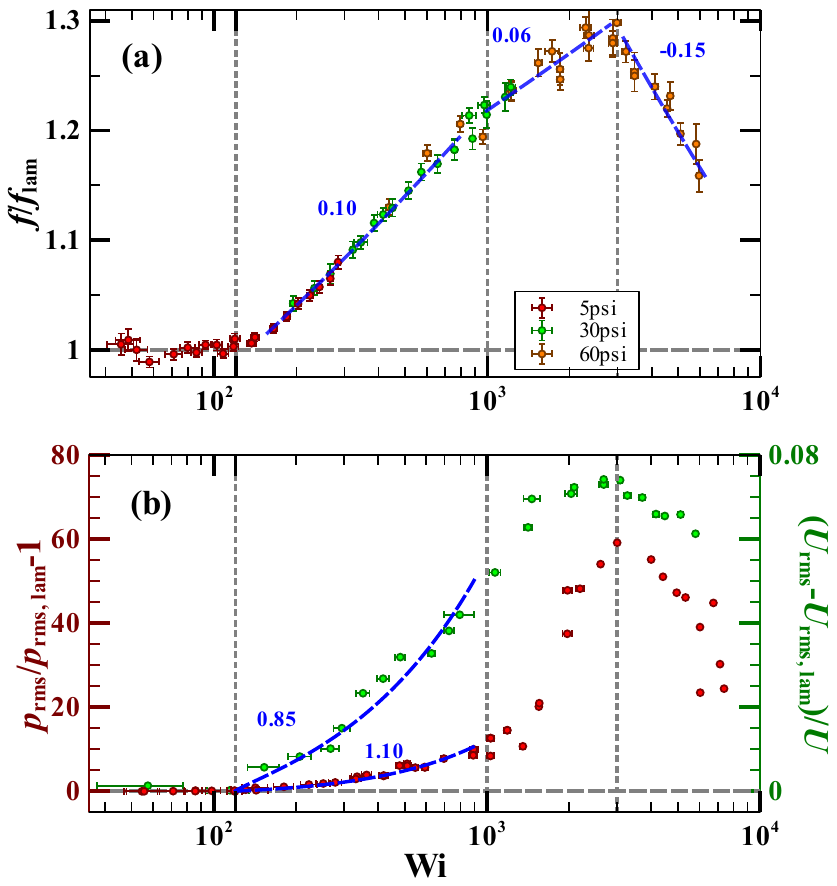}
\caption{
(a) The friction factor ($f/f_{\mathrm{lam}}$) versus $Wi$ in lin-log scales. 
Two individual pressure drop measurements are conducted at different positions but with the same gap: from $l/h$=185 to 725 and from 225 to 765, which are plotted together. Three differential pressure transducers in three ranges are used: 5 (in red), 30 (in green), and 60 (in orange) psi. 
(b) Normalized pressure ($p_{rms}/p_{rms,lam}-1$) and velocity  ($(U_{rms}-U_{rms,lam})/U$) fluctuations versus $Wi$ in lin-log scales. 
}
\label{fig:2_transition}
\end{figure}


In Fig.\ref{fig:2_transition} (a) we present the measurements of the frictional drag as a function of $Wi$ in high resolution presentation, where the friction factor, $f/f_{\mathrm{lam}}$, is calculated as $f=2D_h\Delta P/\rho u_{\mathrm{m}}^2L_p$ and normalized by the laminar one $f_{lam}\sim Re^{-1}$. Here $D_h=2wh/(w+h)=0.875$ mm is the hydraulic length, $L_p=270$ mm is the distance between two pressure measurement locations, and $\Delta P$ is the pressure difference on the length $L_p$. In Fig.\ref{fig:2_transition} (b), we also measure pressure and velocity fluctuations to additionally characterize $Wi_c$ and different flow regimes. 

Four flow regimes are identified by different dependencies of $f/f_{\mathrm{lam}}$ on $Wi$ (Fig.\ref{fig:2_transition} (a)). In a laminar flow the friction factor is independent on $Wi$ and equal unity up to the elastic instability onset at $Wi_c=120\pm10$. First, $f/f_{\mathrm{lam}}$ enhances with $Wi$ as a power-law with an exponent $0.10\pm0.03$, followed by ET where it grows slower with the exponent $0.06\pm0.02$. And finally in drag reduction (DR) regime, the friction factor decays with the exponent $-0.15\pm0.05$. Thus, the dependence of $f/f_{\mathrm{lam}}$ on $Wi$ reveals four flow regimes separated by three grey dashed lines in Fig.\ref{fig:2_transition}.

Pressure fluctuations are normalized by pressure fluctuations in laminar flow, as $p_{\mathrm{rms}}/p_{\mathrm{rms,lam}}-1$, as well as velocity fluctuations reduced by velocity fluctuations at laminar regime normalized by mean flow velocity, as $(U_{\mathrm{rms}}-U_{\mathrm{rms,lam}})/U$ (Fig.\ref{fig:2_transition}(b)). The velocity fluctuations are measured at the center of middle plane via PIV. Analogously to the friction factor, both the normalized pressure and velocity fluctuations versus $Wi$ disclose four flow regimes as well.
$p_{\mathrm{rms}}/p_{\mathrm{rms,lam}}-1$ grows algebraically with an exponent of $0.85\pm0.1$ and $(U_{\mathrm{rms}}-U_{\mathrm{rms,lam}})/U$ with $1.10\pm0.1$ above transition. These exponents differ from 0.5 indicating that the elastic instability is not a linear normal mode bifurcation \cite{drazin2004hydrodynamic}. 

\begin{figure}
\includegraphics{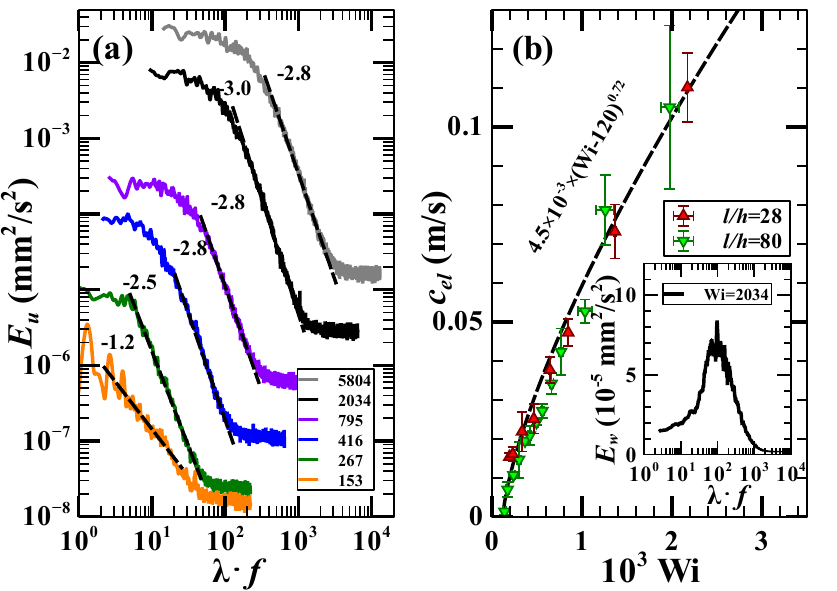}
\caption{(a) Energy spectrum of stream-wise velocity ($E_u$) versus frequency normalized by $\lambda$ at $l/h$=330 in log-log scales. The velocity for each moment is measured at channel center.
(b) Elastic wave speed ($c_{el}$) versus $Wi$ at $l/h$=28 and 80 in lin-lin scales.
The inset presents a peak of elastic wave in span-wise velocity energy spectrum ($E_{w}$) at $Wi=2034$ at $l/h$=330 in lin-log scales.
}
\label{fig:3_spectra}
\end{figure}
Fig.\ref{fig:3_spectra}(a) presents energy spectra ($E_u$) of stream-wise velocity versus normalized frequency ($\lambda\cdot f$) at $l/h$=330, far downstream from the inlet, and the decay slopes are algebraically fitted. The slope exponents decrease from -1.2 to -2.8 in the transition regime, then further to -3.0 in ET, and finally increases back to -2.8 in DR. The chaotic continuous modes of power spectra just above the instability are  another strong evidence of the non-modal nature of the elastically driven instability \cite{hariharan2021localized,jovanovic2010transient,jovanovic2011nonmodal}.

\begin{figure*}
\includegraphics{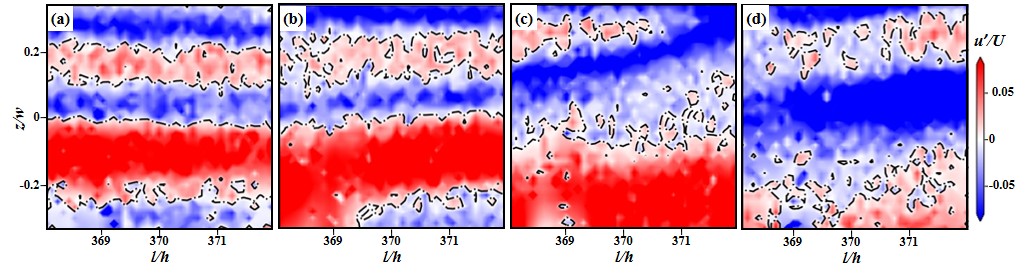}
\caption{Streak interface dynamics at $l/h$=330 and $Wi$=2030 in ET during a single cycle. The time-averaged stream-wise velocity profile is subtracted from the instant stream-wise velocity and normalized by time-averaged velocity $u'/ U$. The black dashed line separates the low profile (blue) and high profile (red) streak interface. The moment $t^*=tf_{el}$ of each image from left to right is 0.20, 0.38, 0.65, and 0.92. The interfaces of streaks are gradually interfered and break into random at the end of the cycle without any secondary instability.
}
\label{fig:4_streaks}
\end{figure*}
\begin{figure}
\includegraphics{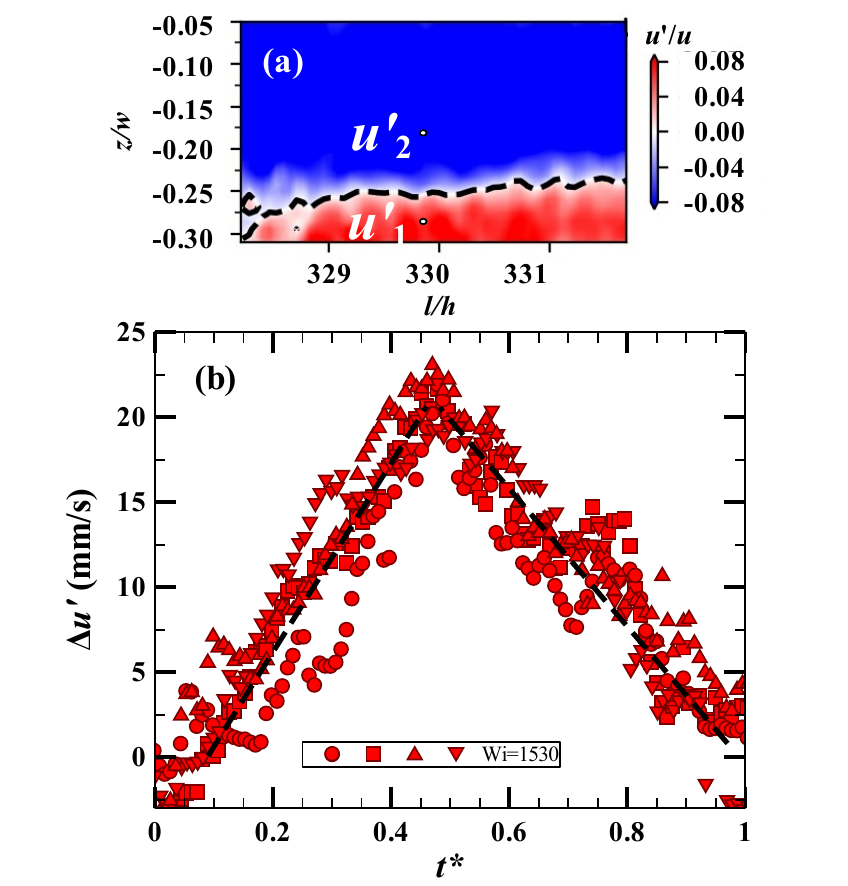}
\caption{Temporal dynamics of streaks at $l/h$=330 at $Wi$=1530. (a) Span-wise-separated two points for calculation of $\Delta u'=u'_2-u'_1$.
(b) Four runs of $\Delta u$ for one cycle normalized by the period of elastic wave.
}
\label{fig:5_streaks2}
\end{figure}
Using span-wise velocity energy spectra ($E_{w}$) in lin-log scales (see an example in the inset in Fig.\ref{fig:3_spectra}(b))
, we obtain the dependence of the elastic wave frequency on $Wi$, the first indication of the observation of elastic waves. The main proof of the elastic wave detection to be compared with our early observations \cite{varshney2019elastic,jha2020universal,jha2021elastically,shnapp2021nonmodal} is the result on the elastic wave speed ($c_{el}$) scaling with $(Wi-Wi_c)^\delta$. Fig.\ref{fig:3_spectra}(b) plots $c_{el}$ vs $Wi$ in lin-lin scales obtained by cross-correlation of velocity along stream-wise direction.
By fitting the curve with $c_{el}=A\times(Wi-Wi_{c})^{\delta}$, we obtain $A=4.5\pm0.5\times 10^{-3}$ m/s and $\delta=0.72\pm0.02$, which are the same as found in earlier flows: flows between two widely separated obstacles \cite{varshney2019elastic}, and a straight channel with an array of obstacles at the inlet to excite strong perturbations \cite{jha2020universal,jha2021elastically}. On the other hand, span-wise propagating elastic waves discovered in a straight channel with very weak perturbations generated by a small cavity in the top plate exhibit the coefficient $A$ of about three orders of magnitude less and the same scaling exponent that maybe attributed to the strong span-wise confinement 
\cite{shnapp2021nonmodal}. This suggests that the stream-wise-propagating elastic wave scaling relation with $Wi-Wi_c$ could be universal. 

To gain better insights into coherent states, we broaden the PIV window to the whole channel 
and at the stream-wise length $ \Delta l/h=4$  to examine probable coherent states in this channel flow. 
As illustrated in Fig.\ref{fig:4_streaks} from left to right in a single elastic wave cycle, at the beginning 
streaks first are self-organized with slightly perturbed interface that further is perturbed more and finally destroyed leading to a random flow at the end of the cycle. Then the next cycle starts. The streak dynamics is synchronized by elastic waves period. To prove experimentally the periodicity of a cycled self-sustained process with the period of the elastic waves we utilize the approach developed by us early to study the interface dynamics \cite{jha2020universal,jha2021elastically} by examining a temporal evolution of $\Delta u'=u'_2-u'_1$, the velocity fluctuation difference at two specific points across the interface,
as shown in Fig. \ref{fig:5_streaks2}.
During each cycle, $\Delta u'$ displays a non-monotonic temporal variation with the same peak values $\Delta u'_{max}$ and cycle period. 
It should be pointed out that no secondary instability of the streaks such as the Kelvin-Helmholtz-like elastic instability is observed, which we attribute to lower intensity of the elastic waves compared to that found in Ref. \cite{jha2021elastically}.


To conclude, first of all, the main result of this study is resolving the recent puzzle whether strong prearranged perturbations are necessary to get an elastic instability in parallel shear visco-elastic flows. Indeed, we investigate in a straight planar channel visco-elastic flow the elastically induced instability. In contrast to previously published works, we remove obstacles in the channel, while the non-smoothed inlet and six holes in the top plate along the channel provide perturbations that lead to the intrinsic non-modal instability at $Wi_c=120$.
We find and study four flow regimes: laminar, transitional, ET, and DR, which are characterized by measuring the friction factor, pressure and velocity fluctuations, and stream-wise velocity power spectra scaling. The velocity of stream-wise-propagating elastic wave shows the universal scaling relation with $Wi-Wi_c$. Using the approach to quantify a cycling period of streak dynamics \cite{jha2020universal,jha2021elastically}, we show that the period of the streak cycling dynamics is equal to the elastic wave period.  

The remarkable and surprising finding is the existence of the streaks and elastic waves over the entire channel length up to $l/h$=980, in a sharp contrast with the case of the strong prearranged perturbations at the inlet \cite{jha2020universal,jha2021elastically}, where the CSs and elastic waves retain only in the short range of the channel from $l/h=36$ up to $l/h\approx200$. Such distinct difference between two types of external perturbations may be explained by the following arguments. Finite-size perturbations are necessary to excite an intrinsic elastic instability in a channel flow. The presented above results of various measurements show that the instability occurs due to non-normal mode bifurcation, which sensitively depends on initial conditions \cite{schmid2007nonmodal} including disturbances. It is revealed in different CSs observed in two cases: only streaks for small perturbations versus rolls and streaks for strong ones. Moreover, in the current case, besides the perturbations from inlet generated stream-wise elastic waves, six holes along the channel further support the elastic waves by pumping energy. It is clearly characterized by measurements of their attenuation along the channel, of which  details will be published elsewhere. On other hand, the current case has a reduced intensity of elastic waves that 
results in the absence of the secondary instability in streaks, namely the Kelvin-Helmholtz-like elastic instability discovered recently in a channel viscoelastic flow with strong perturbations \cite{jha2021elastically}.  


We are grateful to Guy Han and Rostyslav Baron for their help with the experimental setup. This work was partially supported by grant from the Israel Scientific Foundation (ISF, grant \#784/19.)

\nocite{*}
\bibliographystyle{unsrt}
\bibliography{main}
\end{document}